\documentclass[conference]{IEEEtran}
\usepackage{cite}
\usepackage{amsmath,amssymb,amsfonts}
\usepackage{algorithmic}
\usepackage{graphicx}
\usepackage{svg}
\usepackage{textcomp}
\usepackage{subcaption}
\usepackage{xcolor}
\usepackage{booktabs}
\usepackage{hyperref}

\usepackage{tikz}
\usepackage{eso-pic}
\usepackage{everypage}

\def\BibTeX{{\rm B\kern-.05em{\sc i\kern-.025em b}\kern-.08em
    T\kern-.1667em\lower.7ex\hbox{E}\kern-.125emX}}

\AddEverypageHook{%
  \ifnum\value{page}=1 %
    \begin{tikzpicture}[remember picture,overlay]
      \node at ([yshift=-30pt]current page.north) {%
        \parbox{0.85\paperwidth}{%
          \centering
          \color{gray!60}
          \fontfamily{cmr}\fontseries{m}\fontsize{12pt}{12pt}\selectfont
          This paper has been accepted for publication in 2024 IEEE Biomedical Circuits and Systems Conference (BioCAS) in Xi'an, China.
        }%
      };
    \end{tikzpicture}%
  \fi
}

\AddEverypageHook{%
  \ifnum\value{page}=1 %
    \begin{tikzpicture}[remember picture,overlay]
      \node at ([yshift=40pt]current page.south) {%
        \parbox{0.85\paperwidth}{%
          \centering
          \color{gray!60}
          \fontfamily{cmr}\fontseries{m}\fontsize{9pt}{9pt}\selectfont
          \textcopyright\ 2024 IEEE. Personal use of this material is permitted. Permission from IEEE must be
          obtained for all other uses, in any current or future media, including
          reprinting/republishing this material for advertising or promotional purposes, creating new
          collective works, for resale or redistribution to servers or lists, or reuse of any copyrighted
          component of this work in other works.
        }%
      };
    \end{tikzpicture}%
  \fi
}

\begin{document}

\title{Leveraging  Recurrent Neural Networks for Predicting Motor Movements from Primate Motor Cortex Neural Recordings}

\author{
\IEEEauthorblockN{Yuanxi Wang, Zuowen Wang, Shih-Chii Liu}
\IEEEauthorblockA{\textit{Institute of Neuroinformatics} \\
\textit{University of Zurich and ETH Zurich}\\
Zurich, Switzerland \\
yuanxwang@ethz.ch, \{zuowen, shih\}@ini.uzh.ch}
}

\maketitle

\begin{abstract}

This paper presents an efficient deep learning solution for decoding motor movements from neural recordings in non-human primates. An Autoencoder Gated Recurrent Unit (AEGRU) model was adopted as the model architecture for this task. The autoencoder
is only used during the training stage to achieve better generalization. Together with the preprocessing techniques, our model achieved 0.71 $R^2$ score, surpassing the baseline models in Neurobench and is ranked first for $R^2$ in the IEEE BioCAS 2024 Grand Challenge on Neural Decoding. Model pruning is also applied leading to a reduction of 41.4\% of the multiply-accumulate (MAC) operations with little change in the $R^2$ score compared to the unpruned model.

\end{abstract}


\section{Introduction}

Millions of people around the world suffer from a form of body paralysis. 
For those who suffer injury or degradation of the peripheral nervous system or spinal cord but still retain intact brain function,  intracortical brain-machine interfaces (iBMIs) can be a promising solution for partial restoration of the functionality of their limbs.
Neural signals from the motor cortex and premotor cortex of patients can be recorded using multi-electrode recording arrays like the Utah micro-electrode array~\cite{maynard1997utah}. Decoding these neural signals can help one to uncover the underlying movement intention of the patient, and serves for the downstream tasks such as prostheses controlling~\cite{gao2020recurrent}. 

With the availability of more neural recording datasets, researchers can develop new  decoding algorithms 
to determine if they can give accurate predictions of the input stimuli or kinematic states from neural recordings~\cite{zhou2023annvssnncase,ahmadi2021robust,hadorn2022fast, zhou2023ann}. Non-linear decoders particularly those based on deep learning algorithms are also widely assessed on these recordings. 
They consistently report better task accuracy than a linear Kalman filter~\cite{shaikh2019towards, hosman2019bci}. 
For the deep learning models used for neural decoding, 
features such as multi-unit activity (MUA)~\cite{shaikh2019towards, hosman2019bci}, entire spiking activity (ESA)~\cite{ahmadi2021robust}, and local field potential (LFP)~\cite{ahmadi2019decoding} are usually first extracted from the raw neural data and used as input to the network. Data preprocessing steps can help highlight the most relevant neural signals, and improve the signal-to-noise ratio, ultimately enhancing the accuracy and reliability of the decoder. 

In this study, we present a neural network-based solution that is both computationally efficient and uses a low memory footprint for decoding the primate reaching motion speed from intracortical neural recordings. 
The network is trained on recordings of 
the primate reaching dataset was presented as part of the IEEE BioCAS 2024 Grand Challenge on Neural Decoding. \textcolor{black}{The results of our experiments show that the time duration of the input feature to our model plays a crucial role for the accuracy in the task of finger velocity prediction, emphasizing the importance of data preprocessing of neural decoding.} We benchmark our method against baselines using Neurobench~\cite{neurobench_yik2023neurobench}. 


\section{Methods}
This section describes the dataset and metrics used in this study, then introduces our AEGRU model architecture, training methods, and methods for reducing the memory footprint and computes for the final model. 
\subsection{Dataset}
\label{sec:dataset}
We use the six recordings selected in Neurobench~\cite{neurobench_yik2023neurobench} from a bigger dataset released by~\cite{dataset_o2017nonhuman}. 
This dataset contains intracortical recordings from microelectrode arrays implanted in two non-human primates (NHP) named Indy and Loco, while they performed a reaching task to a target location.
These reaches were done by controlling a cursor in the workspace with their fingertips. Once the target was reached, a new target appeared without gaps or pre-movement delay intervals. 
Spikes in the brain activity were detected by means of the threshold-crossing technique after the baseline drift was removed. Spike data were finally sampled at $250 Hz$, the same rate used to sample the finger velocity of the NHP.
From the six recordings in the Challenge dataset, 3 recordings came from Indy and the other 3 recordings came from Loco.
They are: \textit{indy\_20160622\_01, indy\_20160630\_01, indy\_20170131\_02, loco\_20170131\_02, loco\_20170215\_02, loco\_20170301\_05}. Recordings from Indy and Loco came from 96 and 192 input channels of the microelectrode arrays respectively.

\subsection{Metrics}
The decoding performance of the models on the Grand Challenge dataset 
is measured by the Neurobench harness~\cite{neurobench_yik2023neurobench}.  
They consist of the following three metrics: $R^2$ accuracy, memory footprint, and computational complexity measured by the effective multiply-accumulate (MAC) and accumulate (AC) operations.
\subsubsection{$R^2$ accuracy}
The alignment between the ground truth and predicted velocities is quantified by the $R^2$ score:

\begin{equation}
R_x^2 = 1 - \frac{\|\boldsymbol{v_x}-\boldsymbol{\hat{v}_x}\|_2^2}{\|\boldsymbol{v_x}-\overline{v}_x\|_2^2}, \hspace{0.3cm}
R_y^2 = 1 - \frac{\|\boldsymbol{v_y}-\boldsymbol{\hat{v}_y}\|_2^2}{\|\boldsymbol{v_y}-\overline{v}_y\|_2^2}
\label{eq:r1}
\end{equation}
where $\|\cdot\|_2^2$ denotes the squared of $L_2$ norm, $\boldsymbol{v_x}, \boldsymbol{v_y}$ vectors describe the x and y-components of the actual velocity over time, $\boldsymbol{\hat{v}_x}, \boldsymbol{\hat{v}_y}$ are the vectors of predicted velocity components over time, and $\overline{v}_x, \overline{v}_y$ are scalars of the average x- and y- components of velocity. Minus $\overline{v}_x$ or $\overline{v}_y$ is broadcasted on every timestep of $\boldsymbol{v_x}$ and $\boldsymbol{v_y}$ respectively. The final $R^2$ is averaged over $R^2_x$ and $R^2_y$  as described in Eq.~\ref{eq:r1}.

\subsubsection{Memory footprint}

The memory footprint measurement sums up the memory needed for storing the model parameters and activations at their respective numerical precision. In Neurobench memory footprint does not exclude pruned weights. 

\subsubsection{Computational complexity}

The computational complexity is calculated from the average number of effective MAC operations and AC operations per inference pass through the model. It excludes zero activations and zero connections.

The number of ACs is used to indicate the computation cost of the spiking neural network (SNN) layer, i.e. in the Neurobench baseline models, SNN3D and SNN\_streaming. Although an AC operation uses less computational resources than a MAC operation, for simplicity, we choose to use their sum as a metric 
because the number of ACs is consistently much lower than the number of MACs.

\subsection{Model architecture}
\label{method:arch}

The basic building blocks of the AEGRU model are described next starting with the GRU model and extending to the autoencoder.

\begin{table}[!tb]
\caption{Notation list}
\begin{center}
\begin{tabular}{lp{6cm}}
    \toprule
    \textbf{Notation} & \textbf{Description}\\
    \midrule
    $N$ & Number of GRU time steps (sequence length)\\
    $WS$ & Number of sample points in a window for summation\\
    $C_i$ & Number of multi-electrode input  channels\\
    $C_f$ & Number of latent layer features\\
    $C_h$ & Number of GRU hidden layer features\\
    $C_\sigma$ & Number of hidden layer features for latent factor variance prediction \\
    $x, x'$ & Input and preprocessed input  \\
    $r$ & Reconstructed firing rates (a.u.)  \\
    $h, h_0$ & hidden state and the initial hidden state of GRU layer  \\
    $f, \mu^f, \sigma^f$ & Latent factors and their mean and standard deviation \\
    
    \bottomrule
    \end{tabular}
    \vspace{-0.5cm}
\label{tab:notation}
\end{center}
\end{table}

\subsubsection{GRU}
The proposed AEGRU model is based on the Gated Recurrent Unit (GRU)~\cite{gru_cho2014learning}, a basic building block of gated Recurrent Neural Networks (RNNs) proposed to mitigate the issue of either vanishing or exploding gradients during training and useful previously for processing temporal sequences such as time series biomarker processing
~\cite{wang2022person} due to its capability of streamlining the data processing.
The GRU update equations are shown below:
\begin{equation}
    r(t)=\sigma (W_{ir}x(t)+W_{hr}h(t-1))
\end{equation}
\begin{equation}
    u(t)=\sigma (W_{iu}x(t)+W_{hu}h(t-1))
\end{equation}
\begin{equation}
    c(t)=\tanh (W_{ic}x(t)+r(t)\odot W_{hc}h(t-1)
\end{equation}
\begin{equation}
    h(t)=(1-u(t)\odot c(t)+u(t)\odot h(t-1)
\end{equation}
where $x(t)$ is the current input, $r(t)$ the current reset gate, $u(t)$ the current update gate, $c(t)$ the candidate hidden state, and $h(t-1), h(t)$ the previous and current updated hidden states.  $W$s include both weights and biases of the GRU layer, $\sigma$ is sigmoid activation function, and $\odot$ is Hadamard product.

\subsubsection{Autoencoder}

An autoencoder (AE) is employed during the training process. The AE consists of two components - an encoder and a decoder. The encoder maps the input $x$ to a latent space $f$ and the decoder reconstructs the input $\hat{x}$ based on $f$. The training loss function of the AE includes a term that compares the difference between the input and the reconstructed output. 

\subsubsection {AEGRU inference architecture}
Fig.~\ref{fig:model_arch} shows the architecture consisting of a GRU layer of 32 neurons as the core, flanked by two fully connected (FC) layers. The FC layer before the GRU layer has 32 neurons while the FC layer after the GRU layer has 2 neurons. Our inference architecture has a memory footprint of 45.5 kB and thus could be deployed on edge devices.
The network inference formulation is as follows: 
\begin{equation}
f=\text{FC}_{\text{upstream}}(x')
\end{equation}
\begin{equation}
\label{eq:gru}
h=\text{GRU}(f, h_0)
\end{equation}
\begin{equation}
\label{eq:fc_down}
v=\text{FC}_{\text{downstream}}(h)
\end{equation}
where $x' \in \mathbb{R}^{N\times C_i}$ is the preprocessed input, $f \in \mathbb{R}^{N\times C_f}$ the latent factor, $h, h_0 \in \mathbb{R}^{C_h}$ the GRU hidden state, and $v=(v_x, v_y)  \in \mathbb{R}^2$. Other variables are defined in Tab.~\ref{tab:notation}.

\subsubsection {AEGRU training}
\label{methods:model:train}

In addition to the backbone FC-GRU-FC components for inference,  an auxiliary branch is included and used specifically  only during training. This branch bifurcates from the output of the upstream FC layer. It is used to reconstruct the firing rate of the input spikes computed using the spike count within a window. This auxiliary branch combined with the upstream FC layer, forms an autoencoder (AE) to better align the intermediate latent factors with the underlying features of the input spike counts. The mean of the latent factor $f$, $\mu_f$, as generated by the upstream FC layer, serves as the input to the GRU layer in the forward phase. The variance of $f$, $\sigma^2_f$, is generated by the 2-layer FC network (FC1 and FC2). The sampled $f$ from the $\mu_f$ and $\sigma^2_f$, is passed to the GRU layer. 
On the side of the auxiliary branch, $f$ is used to reconstruct the firing rate $r$ using FC3 followed by an exponential transformation. The forward pass for auxiliary training is formulated as below:

\begin{figure}[!tb]
    \centering
    \includegraphics[width=0.8\linewidth, trim=0cm 1.5cm 0cm 0.5cm, clip]{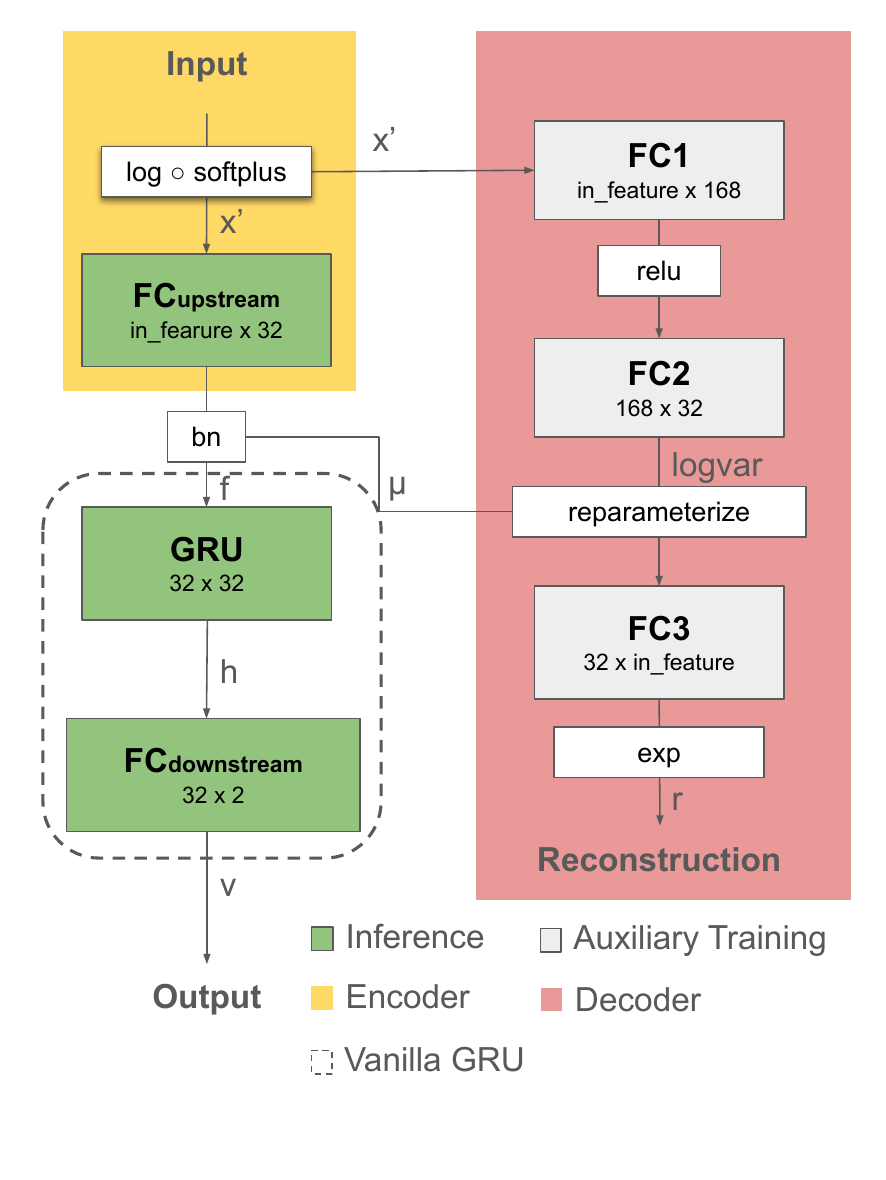}
    \caption{Our AEGRU network architecture. Green blocks represent the inference pass, and the grey blocks compose the auxiliary branch used during training. The weight dimensions are indicated in the blocks. ‘bn'=batch normalization.}
    \label{fig:model_arch}
    \vspace{-0.5cm}
\end{figure}

\begin{equation}
\mu_f = \text{FC}_{\text{upstream}}(x')
\end{equation}
\begin{equation}
\log \sigma_f^2 = \text{FC2}(\text{ReLU}(\text{FC1}(x')))
\end{equation}
\begin{equation}
f\ \sim \  \text{Gaussian}(f| \mu_f, \sigma_f^2)
\end{equation}
\begin{equation}
r = \exp(\text{FC3}(f))
\end{equation}


 The training loss function $L = w_vL_v + w_xL_x$ is comprised of two terms: $L_v$, the mean square error (MSE) between the velocity prediction $\hat{v}$ and ground truth $v$; and $L_x$, the Poisson negative log-likelihood (Poisson NLL) between the reconstructed firing rate, $r$, and the input $x$ as used for auxiliary training:
\begin{equation}
L_v = \frac{1}{t} \sum_t ||\hat{v_t}-v_t||^2_2
\end{equation}
\begin{equation}
L_x = -\frac{1}{t}\sum_t \log(\text{Poisson}(x_t|r_t))
\end{equation}

\subsection{Data preprocessing}
\label{methods:model:preprocess}
\subsubsection{Neural activity representation} 

\begin{figure}[!t]
\centering
        \centering
        \includegraphics[width=0.9\linewidth, trim=0cm 7.2cm 0cm 2.5cm, clip]{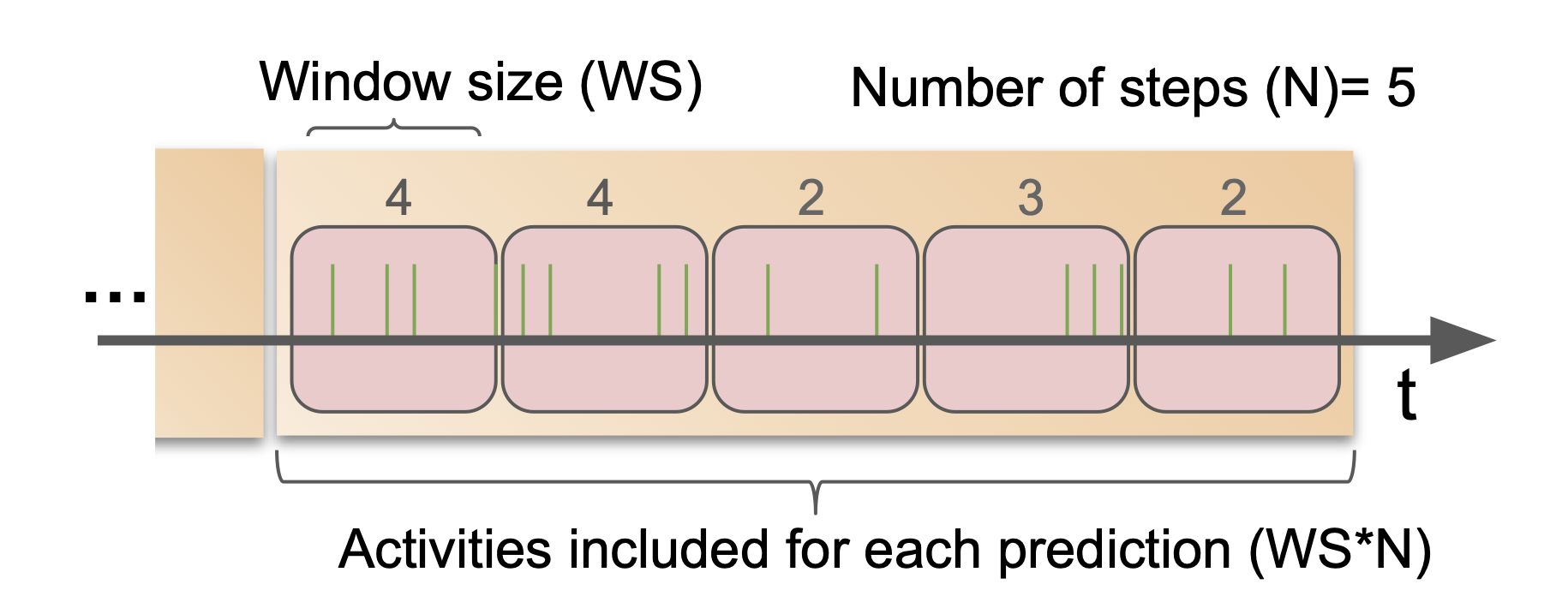}
        \caption{Data pre-processing pipeline. 
        Each input represents $WS \times N \times 4$ ms of neural activity.}
        \label{fig:preproc_pipeline}
\end{figure}

\textcolor{black}{
Following the raw data processing and sub-window method in~\cite{zhou2023annvssnncase}, we bin the data samples into windows of size $WS$. Note that each $WS$ represents $WS \times 4$ ms of spike data. As illustrated in Fig. \ref{fig:preproc_pipeline}, the input feature is defined by first choosing $WS$ number of data samples, then the data samples (firing rates) in each window are summed and $N$ number of steps of such windows are taken to represent the neural activity within the time interval of $WS \times N \times 4$ ms.} 


\textcolor{black}{For every prediction, we use the latest $WS\times N$ consecutive data samples which are then collapsed in $N$-dimensions for each channel after summation within the sub-windows spanned $WS$ samples, as shown in Fig.~\ref{fig:preproc_pipeline}. For the next prediction, we take a stride of 4ms, i.e. 1 data sample. The final input is of dimension $\mathbb{R}^{N\times C_i}$. }

\subsubsection{Softplus and logarithm} 


A softplus function is applied to the spike count data so that all values are positive.
Assuming that the data is Poisson distributed, 
a logarithm is used to transform the data into a Gaussian-like distribution. 
This introduced bias helps to  uncover the underlying Gaussian distributed latent factors through a simple linear transformation.
\begin{equation}
\text{Softplus}(x) = \log(1+\exp(x))
\end{equation}
\begin{equation}
x' = \log(\text{Softplus}(x))
\end{equation}

\subsection{Training details}
Separate models are trained for each of the six recordings. The data split is 50\% for the training set; the middle 25\% and the last 25\% form the validation and test sets respectively.  
Models are trained over 50 epochs using the training set. The Adam optimizer is used with a learning rate of 0.001 and a weight decay of 0.001. The cosine annealing learning rate scheduler is used to adjust the learning rate dynamically over the epochs. Fine-tuning is conducted over 10 epochs after weight pruning (see Sec.~\ref{sec:pruning}) using both training and validation sets.


\begin{figure}[!t]
        \includegraphics[width=0.9\linewidth, trim=0.cm 0cm 0.0cm 0cm, clip]{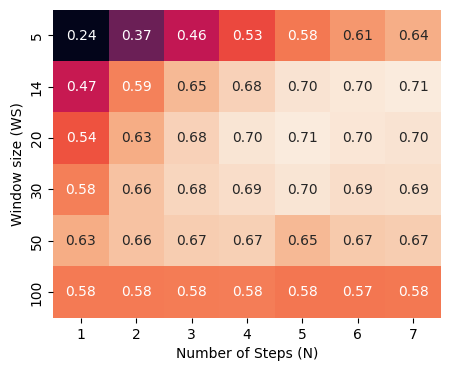}
        \caption{Heat map of mean $R^2$ on the test sets of the six recordings, evaluated across  window sizes ($WS$) and number of steps ($N$).
        }

\label{fig:hyper:ws_n}
\vspace{-0.5cm}
\end{figure}
\subsection{Post-training weight sparsification} 

\subsubsection{Pruning}
\label{sec:pruning}
To reduce the computational complexity,  the model weights were sparsified  
using the L1 unstructured post-training pruning technique. A target sparsity rate of 50$\%$ was specified for both the FC and GRU weights. 
Following pruning, the model was fine tuned over several epochs to restore task accuracy.

\subsubsection{Post-rounding weight quantization} 
The weights were further quantized into an 8-bit fixed-point format, where 1 bit was allocated for the integer part and the remaining 7 bits for the fractional part. 
For each weight, $W$, the quantized weight was achieved using : $W_q = \text{round}(2^{\text{qf}} \cdot W) / 2^{\text{qf}}$ where $\text{qf}$ is the number of fractional bits.

\section{Results}
\label{sec:results}

\begin{figure}[!b]
    \centering
    \includegraphics[width=\linewidth, trim=0.cm 0cm 0.0cm 0.1cm, clip]{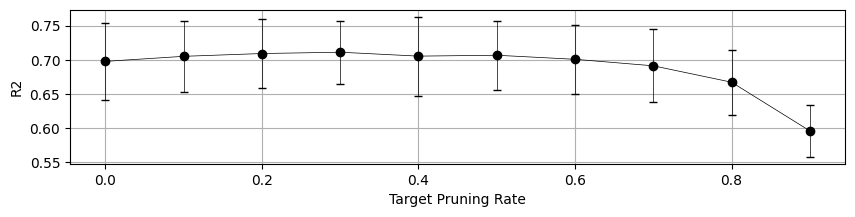}
    \caption{Change in  $R^2$ when the target pruning rate (TPR) is increased. The mean $R^2$ starts to decrease when TPR$> 0.6$. 
    }
    \label{fig:hyper:prune}
\end{figure}

\subsection{Hyperparameter optimization}
\label{sec:hyper}
\subsubsection{Windowing method}

To complete one model execution, the number of data samples
for each channel 
is $WS \times N$ (see Table~\ref{tab:notation} and Fig.~\ref{fig:preproc_pipeline}). Further, $N$ affects the computational complexity, as it sets the number of recurrent steps within the GRU layers. \textcolor{black}{A smaller $WS \times N$ (less data) may cause the model to predict without context of the general kinematics, but a larger $WS \times N$ (longer history of neural activity) may challenge our small model fail to learn.} Thus, there is a trade-off to be optimized between the prediction accuracy and the requirements for memory storage and computation complexity. 

A hyperparameter search was carried out for a range of $WS$ and $N$ settings. 
The mean $R^2$ results in Fig.~\ref{fig:hyper:ws_n} show that 
the $WS=14, N=7$ configuration achieved the highest mean $R^2$ score of $0.709\pm0.04$.  The default configuration, $WS=20, N=5$, had a close score of $0.707\pm0.05$ but used $29.4\%$ fewer MACs than the $WS=14, N=7$ configuration. 

    




\subsubsection{Pruning rate}


Following the pruning method described in Sec.~\ref{sec:pruning}, 
Fig.~\ref{fig:hyper:prune} shows that $R^2$ is maintained even with increasing pruning rate until a target pruning rate (TPR) of 0.6.
For the reported numbers in Table~\ref{tab:results:overview}, the TPR=0.5. The MACs are reduced by 41.4\% at a TPR=0.5 compared to TPR=0. The footprint calculation in Neurobench does not rule out pruned weights thus the footprint size remained the same.

\begin{table}[!t]
\caption{Performance of baseline models and AEGRU. Mean and standard deviation are calculated for all recordings.}
\centering
\begin{tabular}{|l|l|l|l|}
\hline
\textbf{Model} & \textbf{R2}         & \textbf{Footprint (kB)} & \textbf{MACs + ACs (k)} \\ \hline
ANN2D          & 0.58 ± 0.06           & 27.2 ± 6.3               & 5.0 ± 1.2                 \\ \hline
ANN3D          & 0.62 ± 0.05           & 137.8 ± 43.2             & 11.5 ± 3.9                \\ \hline
SNN3D          & 0.63 ± 0.05           & 34.0 ± 9.0               & 38.1 ± 11.6               \\ \hline
SNN Streaming & 0.58 ± 0.06           & 29.2 ± 9.6               & 0.4 ± 0.2                 \\ \hline
\textbf{AEGRU w/o pruning}  &  0.70 ± 0.06 & 45.5 ± 6.3  &  42.8 ± 4.1 \\ \hline
\textbf{AEGRU 50\% pruning}          & 0.71 ± 0.05           & 45.5 ± 6.3               & 25.1 ± 3.3                \\ \hline
\end{tabular}
\label{tab:results:overview}
\end{table}

\begin{table}[!tb]
\caption{Comparison of $R^2$ scores between AEGRU and vanilla GRU. The means and standard deviations are calculated on 5 runs for each recording.}
\centering
\begin{tabular}{|l|l|l|}
\hline
\textbf{Recording} & \textbf{vanilla GRU} &   \textbf{AEGRU} \\ \hline
indy\_20160622\_01 & 0.74 ± 0.006 & 0.76 ± 0.014 \\ \hline
indy\_20160630\_01 & 0.64 ± 0.011 & 0.65 ± 0.008 \\ \hline
indy\_20170131\_02 & 0.75 ± 0.004 & 0.76 ± 0.006 \\ \hline
loco\_20170131\_02 & 0.61 ± 0.017 &  0.68 ± 0.013 \\ \hline
loco\_20170215\_02 & 0.60 ± 0.009 & 0.62 ± 0.014 \\ \hline
loco\_20170301\_05 & 0.66 ± 0.013 & 0.70 ± 0.015 \\ \hline
\textbf{Average} & 0.67 ± 0.005 & 0.70 ± 0.008 \\ \hline
\end{tabular}
\label{tab:results:ablation}
\vspace{-0.5cm}
\end{table}

\subsection{Comparison to baseline models}

Neurobench~\cite{neurobench_yik2023neurobench} provides four baseline models: ANN2D, ANN3D, SNN3D and SNN$\_$streaming on this dataset. These 4 baseline model architectures are further studied in~\cite{zhou2023ann}. 
Table~\ref{tab:results:overview} shows the trade-off between the $R^2$ accuracy and the other 2 metrics 
for our AEGRU model and the  baseline models. The $R^2$ scores of these baseline models come from the Neurobench code harness~\url{https://github.com/NeuroBench/neurobench} with the following $WS$ and $N$ combination: $WS=50, N=1$ for ANN2D, $WS=7, N=7$ for ANN3D and SNN3D, and $WS=1, N=1$ for SNN\_streaming.  The AEGRU model uses the optimized hyperparameter values as described in Sec.~\ref{sec:hyper}, i.e., $WS=20, N=5$ and post-training TPR of 0.5. The mean and standard dedication are calculated for 1 run per recording.   
Overall, our model achieves the highest mean $R^2$ of 0.71. 




In comparison to the vanilla GRU architecture, the AEGRU model showed a higher $R^2$ score for all 6 recordings. The mean and std deviation are computed over 5 runs with different random seeds (see Table~\ref{tab:results:ablation}). The vanilla GRU model comprises only the GRU layer and the downstream FC layer (see Fig.~\ref{fig:model_arch}). 
The upstream FC layer helps to reduce the memory footprint because of the reduced feature dimension to the GRU layer. 
A possible reason for the higher prediction accuracy is that the model generalizes better and is less prone to overfitting. 




\section{Discussion}
 Our AEGRU model achieved a higher $R^2$
 compared to the baseline models through a combination of model design, auxiliary training, and data preprocessing hyperparameters. \textcolor{black}{Furthermore, we found the data preprocessing hyperparameters were vital to the performance. A self-adapting preprocessing hyperparameter method will be studied in future works. } Moreover, the sparsified AEGRU could be deployed on customized hardware designed for exploiting weight sparsity~\cite{gao2022spartus}. 

\section{Acknowledgement}
This project is partially supported by the European Union’s Horizon 2020 research and innovation programme under grant agreement No 899287.

\bibliographystyle{IEEEtran}
\bibliography{ref}

\end{document}